\documentclass[12pt]{article}
\usepackage[margin=1in]{geometry}
\usepackage{amsmath,amssymb}
\usepackage{graphicx}
\usepackage{tikz}
\usetikzlibrary{circuits.logic.US,arrows.meta}
\usepackage{circuitikz}
\usepackage{booktabs}
\usepackage[numbers,sort&compress]{natbib}
\usepackage{hyperref}
\usepackage{url}

\DeclareMathOperator{\eml}{eml}
\DeclareMathOperator{\edl}{edl}
\DeclareMathOperator{\pre}{pre}
\DeclareMathOperator{\suc}{suc}
\DeclareMathOperator{\inv}{inv}
\DeclareMathOperator{\half}{half}
\DeclareMathOperator{\sqr}{sqr}
\DeclareMathOperator{\minus}{minus}
\DeclareMathOperator{\arsinh}{arsinh}
\DeclareMathOperator{\arcosh}{arcosh}
\DeclareMathOperator{\artanh}{artanh}
\DeclareMathOperator{\pow}{pow}
\DeclareMathOperator{\avg}{avg}
\DeclareMathOperator{\hypot}{hypot}

\begin{document}

\title{All elementary functions from a single operator
}

\author{Andrzej Odrzywo{\l}ek\\
\small Institute of Theoretical Physics, Jagiellonian University, 30-348 Krakow, Poland\\
\small E-mail: andrzej.odrzywolek@uj.edu.pl}

\maketitle

\begin{abstract}
A single two-input gate suffices for all of Boolean logic in digital hardware. No comparable primitive has been known for continuous mathematics: computing elementary functions such as $\sin$, $\cos$, $\sqrt{\phantom{x}}$, and $\log$ has always required multiple distinct operations. Here we show that a single binary operator, 
$$\eml(x,y)=\exp(x)-\ln(y),$$ together with the constant~$1$, generates the standard repertoire of a scientific calculator. This includes constants such as $e$, $\pi$, and $i$; arithmetic operations including $+$, $-$, $\times$, $/$, and exponentiation as well as the usual transcendental and algebraic functions. For example, $e^x=\eml(x,1)$, $\ln x=\eml(1,\eml(\eml(1,x),1))$, and likewise for all other operations. That such an operator exists was not anticipated; I found it by systematic exhaustive search and established constructively that it suffices for the concrete scientific-calculator basis.
In EML (Exp-Minus-Log) form, every such expression becomes a binary tree of identical nodes, yielding a grammar as simple as $S\!\to\!1\mid\eml(S,S)$. This uniform structure also enables gradient-based symbolic regression: using EML trees as trainable circuits with standard optimizers (Adam), I demonstrate the feasibility of exact recovery of closed-form elementary functions from numerical data at shallow tree depths up to 4. The same architecture can fit arbitrary data, but when the generating law is elementary, it may recover the exact formula.
\end{abstract}

\clearpage

\subsection*{Summary paragraph}

Elementary functions such as exponentiation, logarithms and trigonometric functions
are the standard vocabulary of STEM education. Each
comes with its own rules and a dedicated button on a scientific calculator;
our derivations rely on many of them simultaneously, even though we know they
are heavily redundant and can be expressed through one another, e.g. $\sin{x} = \cos(x-\pi/2)$, $\sqrt{x}=x^{1/2}$, etc. 
They are the workhorse of quantitative science, appearing in basic and empirical laws and inside the engines of numerical methods like differential equation solvers, integration quadratures and Fourier analysis \cite{NumericalRecipes}.
In digital electronics, a remarkable fact underlies universality: a single
two-input gate, NAND (the Sheffer stroke), suffices to build any Boolean
circuit \cite{Sheffer1913}. Continuous mathematics has lacked such a primitive: calculators must expose many distinct buttons. 
Classical reductions, from logarithm tables \cite{Napier,Briggs} and the slide rule through Euler's formula \cite{Cotes1722} to the exp-log
representation \cite{Liouville1835} with algebraic adjunctions \cite{Ritt1948}, reduced them to a few, but no further.
Despite this, it remains unclear whether this apparent diversity is intrinsic, or whether a smaller generative basis exists.
Here we show that the operator $\mathrm{eml}(x,y) = \exp(x) - \ln(y)$, together
with the constant 1, does exactly that: it reconstructs arithmetic, all
standard elementary transcendental functions, and constants including 
integers, fractions, radicals, $e$, $\pi$ and~$i$.
In simpler terms, a two-button calculator (1,eml) suffices for everything a full scientific
calculator can do. 
Existence of the EML operator reveals that elementary functions are members of a
much simpler class than previously recognized. Every EML expression is a binary
tree of identical nodes, yielding an exceptionally simple grammar:
$S \to 1 \mid \mathrm{eml}(S, S)$, a context-free language that is isomorphic
to well-studied combinatorial objects like full binary trees and Catalan structures. 
Elementary formulas become circuits \cite{Ulmann+2022} composed of identical elements, much like digital hardware built from NAND gates.
This uniform representation provides a complete and regular search space for
continuous symbolic regression \cite{udrescu2020ai,cranmer2020discovering}: parameterized EML trees can be optimized by standard
gradient methods, and when the generating law is elementary, trained weights
can snap to exact closed-form expressions. In effect, a single trainable
architecture gains the potential to discover \cite{liu2025kan} any elementary formula from data.
The EML operator may be the tip of an iceberg. Preliminary searches have
already returned related operators with even stronger properties, including a
ternary variant that requires no distinguished constant.

\clearpage

\subsection*{Significance statement}
Everyone learns many mathematical operations in school: 
fractions, roots, logarithms, and trigonometric functions ($+, -, \times, /, \sqrt{\phantom{x}}, \sin, \cos, \log, \ldots$),
each with its own rules and a dedicated button on a scientific calculator. Higher mathematics reveals that many of these are redundant: for example, trigonometric ones reduce to the complex exponential. How far can this reduction go? We show that it goes all the way: a single operation, $\eml(x,y)$, replaces every one of them. A calculator with just two buttons, EML and the digit~1, can compute everything a full scientific calculator does. This is not a mere mathematical trick. Because one repeatable element suffices, mathematical expressions become uniform circuits, much like electronics built from identical transistors, opening new ways to encoding, evaluating, and discovering formulas across scientific computing.

\clearpage

\section{Introduction}

Single, reusable primitives play a disproportionately large aesthetic and practical role in mathematics, engineering, and even biology. Widely known classical examples include the NAND gate (and its dual, Peirce Arrow, logical NOR) for Boolean 0/1 logic \cite{Sheffer1913,Lukasiewicz1963Elements}, the operational amplifier \cite{6767960} for positive and negative feedback processes, and, more recently, the rectified linear unit (ReLU) ''ramp'' activation function \cite{NairHinton2010ReLU} in deep learning \cite{GoodfellowDL}. 
We also mention Wolfram's single axiom \cite{NKS}, K,S combinators from combinatory logic \cite{Schonfinkel1924,CurryFeys1958}, Interaction Combinators \cite{Lafont1997InteractionCombinators}, and fuzzy versions of the Sheffer stroke \cite{FuzzyLogicSheffer}. Other well-known examples are one-instruction set computers (OISC), e.g.\ SUBLEQ \cite{mazonka2011}, Conway's FRACTRAN \cite{Conway1987} and the Rule~110 cellular automaton \cite{NKS, cook2004Rule110}. 

Whether the existence of a single sufficient operator or element is conceptually crucial is disputed 
\cite{Linsky2011EvolutionOfPrincipiaMathematica}. Nevertheless, its value for understanding, pedagogy, and public communication is substantial.
Indeed, classical first-order logic does not need to single out NAND at all: it works perfectly well with a whole redundant family of connectives (XOR, AND, NOT, \ldots) that we introduce for convenience in applications. But the realization that all of them are definable from a single primitive is a striking and conceptually deep structural fact. A similar kind of realization accompanies the recognition of DNA and RNA \cite{crickwatson1953implications} as a nearly universal information carrier in evolutionary biology.

Sheffer-type elements are rare, and mining them typically requires time, compute, effort, and a bit of luck. They also appear in seemingly distant areas, for example in discrete geometry. The recently discovered solution to the “einstein\footnote{From German \textit{ein Stein}, “one tile,” not after Albert Einstein.} problem”
 had a major impact on the tiling community \cite{Einstein}. The present author already has one such element in his portfolio, namely an igloo construction brick derived from the (2,3,5) spherical M\"obius triangle \cite{DrHabIgloo}.

Elementary functions, for many students epitomized by the dreaded \emph{sine} and \emph{cosine}, play a central role in quantitative reasoning. They can be combined in countably many ways with integers, rational numbers, and mathematical constants such as $\pi$ and $e$, using the four basic arithmetic operations ($+,-,\times, /$) and exponentiation. Elementary functions are at the core of STEM education and at the foundation of modern technological civilization. This includes not only simple formulas or empirical models, but also most numerical algorithms, for example ordinary differential equation solver RKF45 \cite{Fehlberg1970}, integration rules (Gaussian, Tanh-Sinh, \ldots) \cite{TakahashiMori1974,NumericalRecipes}, and the Fast Fourier Transform \cite{CooleyTukey1965}, whose twiddle factors are elementary functions. Finite expressions in elementary functions are supported\footnote{Exponentiation, unlike the four basic arithmetic operations that are almost always built in as primitive operators, has often been treated as a higher-level operation exposed via library functions rather than as a dedicated infix operator.} in virtually all modern programming languages.  While it is generally known that most standard functions and arithmetic operations are heavily redundant (e.g. $\tan{x} = \sin{x}/\cos{x}$, $\sqrt{x} = x^{1/2}$,  $\arsinh{x} = \ln{(x+\sqrt{1+x^2})}$, $x/y = x \times y^{-1}$ etc.), especially in the complex domain (e.g. $\cosh{x} = \cos{(ix)}$, $\sinh{i x} = i \sin{x}$), they have never been regarded as candidates for a single primitive operator. Historically, two major milestones in understanding elementary arithmetic operations from this point of view are (i) the discovery of logarithms and the subsequent creation of tables \cite{Napier,Briggs} followed by slide rule, and (ii) Euler’s formula $e^{i \pi}+1=0$ \cite{Cotes1722}.
Logarithms were introduced to reduce multiplication to addition. The exp-log pair allows them to be expressed in terms of one another:
\begin{equation}
\label{log_tbl}
x \times y = e^{\ln{x} + \ln{y}}, \quad x + y = \ln{\left( e^x \times e^y\right)}.
\end{equation}
Euler’s formula 
\begin{equation}
\label{Euler}
e^{i \varphi}=\cos{\varphi} + i \sin{\varphi}
\end{equation}
shows that, once the imaginary unit $i\!=\!\sqrt{-1}$ is introduced, trigonometric functions can be viewed as another face of '$\exp$' and '$\ln$' \cite{Liouville1835}.  This perspective leads naturally to what scientists call exp--log functions, namely finite expressions built from variables, named constants, arithmetic operations, together with $\exp$ and $\log$, in the spirit of differential and computer algebra \cite{SHACKELL1990611,Richardson:1996:AEF}. In the classical differential-algebraic setting, one often works with a broader notion of elementary function, defined relative to a chosen field of constants and allowing algebraic adjunctions \cite{Ritt1948}, i.e., adjoining roots of polynomial equations (cf. \verb!Root! in Wolfram Language  \cite{WolframMathematica}). That level of generality is not needed here. The present paper takes the ordinary scientific-calculator point of view: start from a concrete list of familiar constants, functions and operations, and ask how far they can be reduced without losing practical functionality. The precise starting list is given later in Table~\ref{Calc4}.

Many reductions inside that list are, of course, classical and well known; what seems not to have been investigated systematically is the endpoint of the process, namely reduction to a single binary operator paired with a single distinguished constant. In particular, the so-called ''broken calculator'' problem \cite{BrokenCalculatorMathForLove} is a popular formulation of computations with a reduced set of available keys or operations. I~used this approach to construct a sequence of increasingly “primitive” yet fully functional calculators with 4, 3, 2, and finally a single operator. This operator is the EML (Exp--Minus--Log),
\begin{equation}
\label{eml}
\eml{(x,y)} = \exp{(x)} - \ln{(y)}.
\end{equation}
Using the EML, a surprisingly simple binary operator, we can express any standard real elementary function in terms of repeated applications of \eqref{eml} to the input variables $x,y,z,\ldots$
and a single distinguished constant, 1. This constant is needed to neutralize the logarithmic term in \eqref{eml} via $\ln{(1)}\!=\!0$. Computations must be done in the complex domain, e.g: generating constants like $i$ and $\pi$ requires evaluating $\ln(-1)$, so $\eml(x,y)$ internally operates over $\mathbb{C}$ using the principal branch.

If we do not need external input variables, e.g. for use in an actual pocket calculator, two buttons are sufficient to retain full functionality: one binary operator, EML, and one terminal symbol, 1. No further reduction of operator count is possible, because at least one binary operator and at least one terminal symbol are required. The existence of such a binary operator, which is itself an elementary function, is somewhat unexpected.

In fact, the EML Sheffer operator \eqref{eml} is as simple as it appears, and in principle the article could end here; its consequences would eventually surface on their own. Nevertheless, it seems worthwhile to explain the methods used in the search for it (Sect.~\ref{sec:methods}), to present the intermediate and final search results (Sect.~\ref{sec:results}), and to outline potential applications (Sect.~\ref{sec:applications}) that are already visible on the horizon. The article concludes with a short summary, possible follow-up directions,  and open questions.

\section{Methods \label{sec:methods}}

I~employed systematic 'ablation' testing to identify minimal operator sets for a fixed scientific-calculator starting list. The list, given in Table~\ref{Calc4}, contains 36 primitives: named constants and input variables, standard unary functions, and binary operations. I iteratively removed one element from this list (constant, function or binary operation) and verified whether the remaining set could still reconstruct original primitives. The Mathematica/Wolfram core language \cite{WolframMathematica} instruction set (Table~\ref{Mma}, 2nd row) served as a reference for an already-optimized and thoroughly tested for almost 40 years minimalist system. Research started out of curiosity whether this system could be reduced further.

\begin{table}
\caption{\label{Calc4} Starting list of constants/variables, functions (unary) and binary operators used to initiate the reduction procedure. This table also fixes the concrete scientific-calculator notion of ``elementary function'' used throughout the paper: finite expressions built from these named symbols. The $\exp{(x)} = e^x$, $\ln{x}$ is natural logarithm, $\inv{(x)} = 1/x$ denotes reciprocal, $\minus{(x)} = -x$ is a sign flip. Square is denoted by $\sqr{(x)} = x^2$ and $\sigma{(x)} = 1/(1+e^{-x})$ is the logistic sigmoid. Trigonometric and hyperbolic functions with their inverses have usual meaning. In addition to four basic arithmetic binary operations we use arbitrary base logarithm $\log_x{y}$, exponentiation $\pow{(x,y)} = x^y$, arithmetic mean $\avg{(x,y)} = (x+y)/2$ and hypotenuse $\hypot{(x,y)} = \sqrt{x^2+y^2}$.}
\begin{tabular}{c| l | r}
Type       & Elements & Count \\ \hline \\
Constants  & $\pi, e, i, -1, 1, 2, x, y$  & 8 \\[6pt]
Functions  & \parbox{0.6\columnwidth}{$\exp, \ln, \inv, \half, \minus, \sqrt{\quad}, \sqr, \sigma$, $\sin, \cos, \tan, \arcsin, \arccos, \arctan$, $\sinh, \cosh, \tanh, \arsinh, \arcosh, \artanh$} & 20 \\[12pt]
Operations & $+, -, \times, /$, $\log, \pow$, $\avg, \hypot$     & 8 \\[6pt]
\hline \\
Total      &          & 36
\end{tabular}
\end{table}

An operator set was deemed ``complete'' if it could reconstruct all primitives from Table~\ref{Calc4}, on the real axis where appropriate. This includes trigonometric functions and their inverses ($\sin$, $\cos$, $\tan$, $\arcsin$, \ldots), hyperbolic functions and their inverses ($\sinh$, $\cosh$, $\tanh$, $\arsinh$, \ldots), algebraic operations ($\sqrt{\cdot}$, reciprocal, \ldots), and fundamental constants ($\pi$, $e$, $i$, -1, 0, 1, 2, \ldots ).

The challenge of executing such a test is illustrated by the following simple exercise from rational functions generation, attributed to \cite{Newman1982ProblemSeminar}. Given only the three operations:  $\suc(x) = x+1$ (successor), 
$\pre(x) = x-1$ (predecessor), and $\inv(x) = 1/x$ (reciprocal/inverse), let us compute negation, i.e.,  $-x$. The non-obvious solution is:
$$
\suc(\inv(\pre(\inv(\suc(\inv(x)))))) = \frac{1}{\frac{1}{\frac{1}{x}+1}-1}+1 = -x.
$$
This exemplifies the formulas expected to be encountered in our search: nested expressions with depth 5-9 that defy intuitive construction. There is no regular method to find them automatically, except for (human-assisted) brute-force search, nowadays also enhanced using AI \cite{Naskrecki2025}.

Direct symbolic verification proved computationally intractable. Typical Kolmogorov\text{-}style complexity $K$ \cite{Kolmogorov1965} is usually $K \lesssim 7$. In practice it was searched up to $K=9$. Here K denotes the length of a Reverse Polish Notation (RPN) calculator program, equivalent to a formula. To speed up the search I developed a hybrid numeric bootstrapping verification approach, working as follows. We test whether a target operation, e.g. hypotenuse $\sqrt{x^2+y^2}$, can be expressed using a given operator set. Instead of symbolic manipulation on $x,y$, I substituted for $x$ and $y$ algebraically independent (not in exp-log class) transcendental constants, such as e.g., $x=\gamma \simeq 0.577216$ (Euler-Mascheroni constant), $y = A \simeq 1.28243$ (Glaisher-Kinkelin constant). 
Then, $\sqrt{\gamma^2+A^2}$ was evaluated numerically, and compared to all expressions generated from $1$ and \eqref{eml}. This method follows from the simple observation that any generally valid formula must also be valid at a single point. Under the Schanuel conjecture \cite{SchanuelTerzo}, coincidental equality between such expressions is vanishingly unlikely, making this a reliable sieve for formula equivalence candidates. In practice, I compute the double-precision numerical value of the operator sought, and then run constant recognition software (''inverse symbolic calculator'') on this result, which returns a candidate formula in the form of RPN calculator code \cite{SymbolicRegressionPackage} or as an expression tree, depending on the variant of the method used. 
The search procedure is heuristic and serves only to exclude evidently false formulas and discover candidate witnesses; independent verification is separate and is given in Supplementary Information (Part II), which provides symbolic checks, numerical cross-validation, and a constructive completeness proof sketch for the Table~\ref{Calc4} class.

In detail, the search procedure worked as follows. Rather than searching for all elementary operations (Table~\ref{Calc4}) directly in pure EML form (which remains computationally infeasible even numerically), I employed iterative bootstrapping:
\begin{enumerate}
\item Start with two lists: 
    \begin{itemize}
        \item set of constants, functions and operators to be verified for robustness, e.g. $S_0=\{1, \eml \}$
        \item set of constants, functions and operators to be computed, e.g., $C_0=\{\pi, e, i, -1, \ldots \exp, \ln, \ldots, +, -, \times, \ldots \}$ cf.~Table~\ref{Calc4}. The list $C_0$ also defines what I mean by 'computing \emph{all} elementary functions'.
    \end{itemize}
\item Search for a expression computing some element from the list $C_i$ using only primitives from the list $S_i$.
\item If one is found, move it from list $C_i$ to list $S_{i+1}$. In the EML example, the first result found is the formula for $e=\eml(1,1)$.
After the first step, the lists become $S_1 = S_0 \cup \{e\} = \{1, \eml, e \}$, $C_1= C_0 \setminus \{e\} = \{\pi, i, -1, \ldots \exp, \ln, \ldots, +, -, \times, \ldots \}$.  
\item Repeat steps 2-3 until all primitives from initial list $C_0$ are reconstructed, i.e. until $C_i$ becomes empty: $C_i= \emptyset$.
\end{enumerate}
The verification procedure, \texttt{VerifyBaseSet}, was implemented using my own Mathematica SymbolicRegression package \cite{SymbolicRegressionPackage}. A version that is three orders of magnitude faster was recently translated by GPT Codex 5.3 into Rust \cite{SymbolicRegressionPackage}, allowing for re-check of EML in seconds, rather than hours.

The ablation process, i.e. running above search with some of the elements from Table~\ref{Calc4} removed,  yielded progressively smaller calculator configurations (named Calc 3, 2, 1, 0 in Table~\ref{Mma}), each requiring different primitives, constants in particular. Some configurations could generate required constants from arbitrary input on their own (via operations like $x-x=0, e^0=1, 0-1=-1, \ldots$ in Calc 2), while others required specific constants (e.g., Calc 1 requires $e$ or $\pi$, Wolfram requires $i$, cf.~footnote in Table~\ref{Mma}). Then the search stalled, and it became evident that the continuous Sheffer operator, if it exists, is not among the familiar named functions. For the final reduction, I began, guided by intuition gathered over many experiments, to enumerate elementary binary functions as candidate single operators, paired with similarly generated constants. Each candidate pair \{ constant, operator\} was tested using the \texttt{VerifyBaseSet} procedure. This, after numerous failures, and a few discarded false-positives, revealed that the $\eml(x,y)$, given by \eqref{eml}, paired with the constant 1, forms a complete basis for reconstructing the elementary functions. Later I realized that EML is not unique, discovering its close cousins: EDL (requiring constant $e$) and a third variant using $-\infty$ as a terminal symbol. Details are the subject of the next section.

\section{Results \label{sec:results}}

\begin{figure}[thb]
    \centering
    \includegraphics[width=\linewidth]{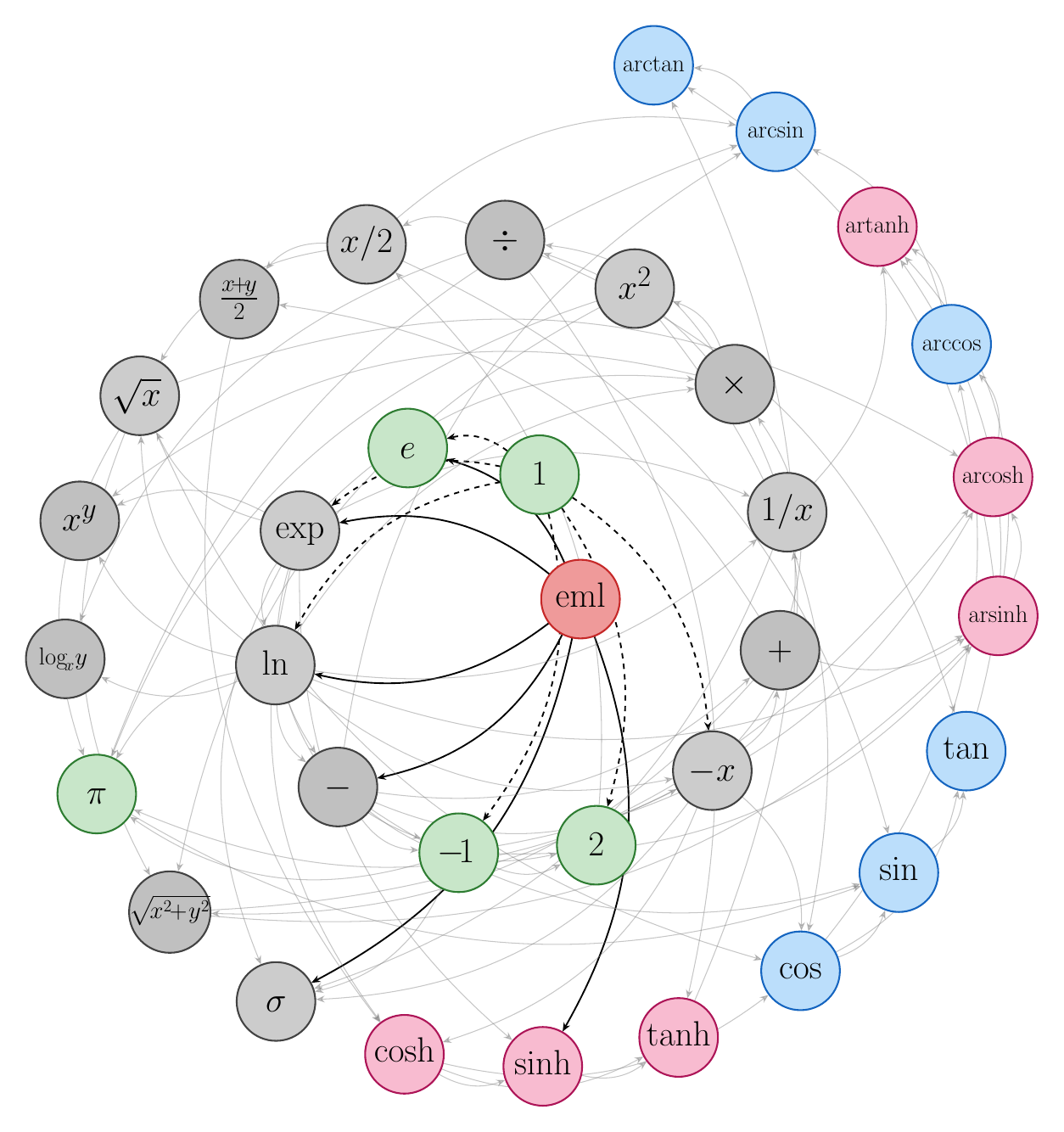} 
    \caption{ \label{fig:phylogenetic} 
	Bootstrapping ``phylogenetic'' tree of the elementary functions found from EML, \eqref{eml}, as LUCA (Last Universal Common Ancestor) and 1. Spiral unwinds according to subsequent found primitives, and arrows show from which elements it was composed of. Those using EML and 1 directly are marked using thick arrows. For full adjacency matrix, and entire discovery chain, see Supplementary Information, Fig.~S1 and Table~S2. 
	}
\end{figure}

\begin{table}
\caption{\label{Mma} Progressive reduction from 36-button scientific calculator from Table~\ref{Calc4} to the EML Sheffer operator. Calc 3, 2 and 0 accept any real constant or variable as input, generating all required values internally. The count-down column tallies all distinct primitives including one input variable; for systems marked ''none'' under Constants, any real value serves as the starting input. Wolfram Mathematica primitives are given for reference.
Last lines are placeholders for speculated undiscovered primitives. The penultimate row is for a binary operator, hypothesized to generate, unlike EML, constants from arbitrary input. The last row is for a unary operator, which retains all good properties of well-tested activation functions in neural nets, while allowing for exact evaluation of elementary functions when combined with standard ''matrix'' arithmetic. }
\begin{tabular}{l|c  c  c  c }
             & Constants              &      Unary             & Binary                 & Count-down   \\ \hline \\
Base-36      & cf.~Table~\ref{Calc4}  & cf.~Table~\ref{Calc4}  & cf.~Table~\ref{Calc4}  & 36 \\[2pt]
Wolfram$^{*}$& $\pi, e, i$           & $\ln$                  & $+, \times, \wedge$    &  7 \\[2pt]
Calc 3       &  none                  & $\exp, \ln, -x, 1/x$   & $+$                    &  6 \\[2pt]
Calc 2       &  none                  & $\exp, \ln$            & $-$                    &  4 \\[2pt]
Calc 1       & $e$ or $\pi$           & none                   & $x^y, \log_{x}{y}$     &  4 \\[2pt]
Calc 0       &  none                  & $\exp$                 & $\log_{x}{y}$          &  3 \\[2pt]
EML          & $1$                    & none                   & $\eml{(x,y)}$          &  3 \\[6pt]
\hline \\[6pt]
 ?           &  none                   & none                  &   ?                   &  2 \\[2pt]
 ?           &  none                   & ?                     &  $+, -, \times$       &  5 \\[2pt]
\end{tabular}

\vspace{4pt}
\noindent{\footnotesize \raggedright $^{*}$The imaginary unit alone is sufficient:
$-1 = i \times i$, $1 = -1 \times (-1)$, $0=-1+1$.
$\pi = -1 \times i \times \ln(-1)$,
$e = (-1)^{(i\pi)^{(-1)}}$.}
\end{table}

Table~\ref{Mma} presents the complete reduction sequence, in historical order. Each configuration represents a complete 'scientific calculator' capable of computing every expression built from the primitives listed in Table~\ref{Calc4}.
Calc 3 (Table~\ref{Mma}, 3rd row) was first to dethrone Wolfram Language primitives set (Table~\ref{Mma}, 2nd row) by retaining negation and reciprocal alongside $\exp$, $\ln$, and addition. This system of 6 primitives is able to generate constants on its own {\em via} $x + (-x) = 0$, etc. 
Calc 2 achieved what Calc 3 does using only $\exp$, $\ln$ and subtraction, while still preserving ability to generate constants on its own. The non-commutative operator (subtraction) proved crucial, as it provides both expression\text{-}tree growth and inversion capabilities.

Calc 1 represents a fundamentally different, top-down approach, using binary exponentiation (a rank-3 hyper-operation) and its inverse (binary logarithm) as base operators instead of lower-rank operations such as addition and subtraction. This configuration works with $e$ or $\pi$ as the terminal constant. Despite extensive searches, no other constant was found for which Calc~1 works.

Calc 0 absorbs the constant $e$ into the $\exp$ function itself, reducing the system to 3 primitives. This configuration strongly suggested that a single binary operator might exist at all, motivating further search.

The final reduction to EML was achieved by recognizing a pattern: all minimal configurations involve pairs of inverse functions (including self-inverses) and non-commutative operations. Testing combinations of inverse functions at the input with asymmetric operations at the output yielded the first continuous sufficient operator. A month later I realized that it has at least two additional cousins: EDL and -EML:
\begin{subequations}
\label{Sheffers}
\begin{align}
\label{eml-1}
\eml(x,y)  &= \exp(x) - \ln(y)  && \text{with constant } 1,\\
\label{edl-e}
\edl(x,y)  &= \exp(x)/\ln(y)    && \text{with constant } e,\\
\label{eml-infty}
-\eml(y,x) &= \ln(x) - \exp(y)  && \text{with constant } -\infty.
\end{align}
\end{subequations}

The first successful discovery search run (cf.~Fig.~\ref{fig:phylogenetic}) can be replicated with this 3-line Mathematica code:
\begin{verbatim}
Import["SymbolicRegression.m"]
EML[x_, y_] := Exp[x] - Log[y]
VerifyBaseSet[{1}, {}, {EML}]
\end{verbatim}
where the package can also be imported directly from the repository \cite{SymbolicRegressionPackage}.
Depending on computer speed, usually in less than an hour, the above procedure re-generates all 36 elementary operations 
from Table~\ref{Calc4}. For example, the natural logarithm becomes:
\begin{equation}
\label{eq:eml-ln}
\ln{(z)} = \eml\left( 1,\eml[ \eml(1,z),1 ] \right),
\end{equation}
and so on.
The resulting EML expressions range from depth 1 (exponential, $e^x = \eml(x,1)$) 
to depth 8 (multiplication), with most basic math functions requiring larger depths (Table~\ref{tbl:leaf_count}). A much faster and more thorough test (multiple real transcendentals, both positive and negative, arbitrary-precision check) is provided by a Rust re-implementation of \texttt{VerifyBaseSet} \cite{SymbolicRegressionPackage}, \texttt{rust\_verify}. 

The exhaustive search served only to discover candidate identities. Their verification is deferred to Supplementary Information, Part~II, which contains symbolic simplification of the full discovery chain (Fig.~\ref{fig:phylogenetic}) in Wolfram Mathematica \cite{WolframMathematica}, independent numerical cross-checks across four implementations (C, NumPy, PyTorch, and mpmath), and a constructive completeness proof sketch.

\section{Usage and applications \label{sec:applications}}

The uniform tree structure of EML expressions suggests several directions for implementation and application.

\subsection{EML compiler \label{subsec:compiler} }

The output of the \texttt{VerifyBaseSet} procedure provides the data (see Fig.~\ref{fig:phylogenetic}) required to reconstruct any primitive or composite elementary expression in terms of EML Sheffer, \eqref{eml-1}. 
I provide a prototype EML compiler, coded in Python, that converts formulas into pure EML form. An EML expression can be evaluated symbolically in Mathematica, or numerically in any IEEE754-compliant language. Pure EML form could also be executed on hardware (or an emulated machine) that has only a~single instruction: the EML itself. In particular, the EML code can be executed on a single instruction stack machine, closely resembling a single-button RPN calculator. Pure-EML form could possibly be implemented efficiently in FPGA or analog circuits.

The simplest input-output example is provided by the $\ln x$ function. After compilation to EML form, we expect to obtain \eqref{eq:eml-ln}, although the simplest possible form is not expected in general. The equivalent RPN code for $\ln$ is a sequence of $K=7$ instructions
$$
{1, 1, x, \eml, 1, \eml, \eml},
$$
or, denoting $\eml \to \text{E}$, as the RPN string \texttt{11xE1EE}. For the expression tree, see Fig.~\ref{fig:trees} (on top).

A few comments on the implementation are required. EML-compiled expressions work on the real axis, both positive and negative, except for a few isolated points, especially at zero and domain endpoints. Internal computations, for trigonometric functions in particular, must be done in the complex domain. 
Because the simplest form of the natural logarithm, \eqref{eq:eml-ln},  obtained from EML, \eqref{eml}, is equivalent to
$$
\ln{z} = e- \log{\left( \frac{e^e}{z} \right )},
$$
using the standard choice of principal branch for complex logarithm, we obtain a jump of $2 \pi i$ for the negative real axis, due to the $1/z$ term. This prevents use of e.g. some standard formulas for $i$, relying on $\ln(-1) = i \pi$, for which we get the wrong sign. A solution working for all real $x \neq 0$ is to redefine the branch for EML itself in such a way that $\ln{z}$ (and everything derived from it, cf.~Fig.~\ref{fig:phylogenetic}) follows standard implementation of principal branch. Another
option, used in EML compiler, is to manually correct $i$ sign.

EML-compiled formulas work flawlessly in symbolic Mathematica and IEEE754 floating-point, e.g. \verb!<math.h>! in C. This is because some formulas internally might rely on the following properties of extended reals:
$$
\ln{0} = -\infty, \quad e^{-\infty} = 0. 
$$
These are properly implemented in Mathematica using symbolic processing, and in floating point using \texttt{inf} and signed zeros. But EML expressions in general do not work 'out of the box' in, e.g.,  pure Python/Julia or numerical Mathematica. In the first case, this is because special floats are trapped and raise errors. However, EML works in \texttt{NumPy} \cite{numpy} and \texttt{PyTorch} \cite{PyTorch}. In Mathematica, we have an automatically extensible range of floats leading to \texttt{Overflow[]}.  
Interestingly, the Lean 4 proof assistant \cite{LEAN4} takes a different approach. Because Lean requires all functions to be total, it assigns the complex logarithm at zero a default ``junk value'' (\texttt{Complex.log 0 = 0}), causing the straightforward formalization of the EML chain to fail.
As a bottom line I stress that all the above difficulties (edge cases) are not much different from those usually encountered in every kind of floating-point or symbolic computation. EML compiler is available for testing under \texttt{EML\_toolkit/EmL\_compiler} subdir of \cite{SymbolicRegressionPackage}; see also Supplementary Information, Sect.~4. \textit{Software and reproducibility}.

\subsection{Elementary functions as binary trees and analog circuits}

Noteworthy, in EML notation, any elementary function expression tree is binary. The context-free grammar is trivial: $S \to 1 \mid \eml{(S,S)}$. 
For functions, input variables are added as additional terminal symbols (e.g. $x$ in the univariate case).
This has many practical advantages over standard methods. 

\begin{figure}
    \centering
    \raggedleft
    \includegraphics[width=\linewidth]{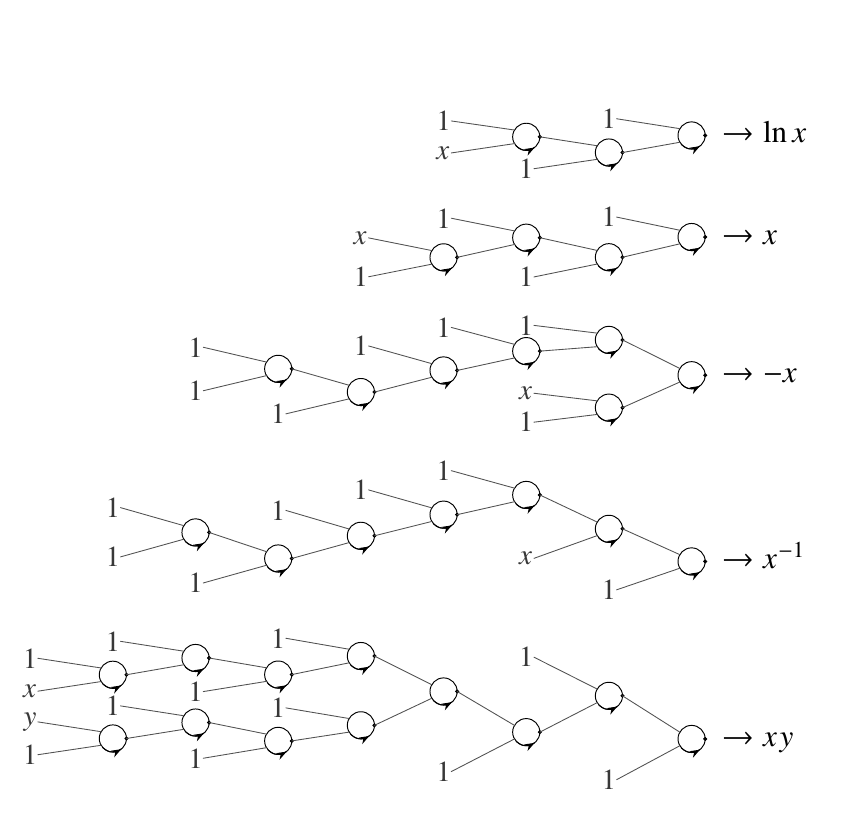}\\
    \caption{
	Examples of binary EML trees equivalent to few important simplest formulas. \label{fig:trees} 
	}
\end{figure}

Some simple examples of tree/circuit representations are shown in Fig.~\ref{fig:trees}. The examples shown are natural logarithm, identity, negation $\minus(x)=-x$, reciprocal $\inv(x)=1/x$, and multiplication. Ability to compute the identity function using an EML tree of depth~4 allows some input variables to be moved down the tree (see next Subsection). Other elementary functions, e.g. trigonometric ones, have trees too large to be shown in print, cf.~Table~\ref{tbl:leaf_count}. 

\begin{table}[htbp]
\centering
\caption{EML Sheffer as a new kind of basic building block for symbolic and analog computations. Dot marks output, and, as \eqref{eml} is non-commutative, arrow determine chirality and operation order. First counterclockwise input after dot is $\exp{x}$, then $\ln{y}$, and dot EML ''perform'' subtraction. In Fig.~\ref{fig:trees} we keep this convention to avoid confusion, but clockwise variant is valid as well. I'm following habits of OpAmp symbol usage here.    
\label{EML-symbol}}
\begin{tabular}{@{}cccc@{}}
\toprule 
NAND gate & Op-Amp & Transistor & EML Sheffer \\[2pt]
\midrule \\[2pt]

\begin{tikzpicture}[circuit logic US, baseline=(N.center), scale=0.9]
  \node[nand gate, draw, logic gate inputs=nn] (N) {};
  \draw (N.input 1) -- ++(-0.4,0);
  \draw (N.input 2) -- ++(-0.4,0);
  \draw (N.output) -- ++(0.4,0);
\end{tikzpicture}
&
\begin{circuitikz}[baseline=(A.center), scale=0.7]
  \draw (0,0) node[op amp, noinv input up] (A) {};
\end{circuitikz}
&
\begin{circuitikz}[baseline=(Q.center), scale=0.7]
  \draw (0,0) node[pnp, xscale=-1] (Q) {};
\end{circuitikz}
&
\begin{tikzpicture}[
    baseline={(0,0)},
    scale=0.48,
    emlwire/.style={draw=black, line width=0.75pt, line cap=round, line join=round},
    emlbody/.style={draw=black, fill=white, line width=0.82pt},
    emlarrow/.style={draw=black, line width=0.82pt, -{Stealth[length=3.4mm,width=2.5mm]}, line cap=round, line join=round}
]
  \draw[emlbody] (0,0) circle [radius=1.34];
  \draw[emlwire] ({2.55*cos(150)},{2.55*sin(150)}) -- ({1.34*cos(150)},{1.34*sin(150)});
  \draw[emlwire] ({2.55*cos(210)},{2.55*sin(210)}) -- ({1.34*cos(210)},{1.34*sin(210)});
  \draw[emlwire] ({1.34*cos(0)},{1.34*sin(0)}) -- (2.18,0);
  \node[circle, fill=black, inner sep=0pt, minimum size=1.55mm] at ({1.34*cos(0)},{1.34*sin(0)}) {};
  \draw[emlwire] ({1.34*cos(220)},{1.34*sin(220)}) arc[start angle=220,end angle=290,radius=1.34];
  \draw[emlarrow] ({1.34*cos(290)},{1.34*sin(290)}) -- ({1.34*cos(308)},{1.34*sin(308)});
\end{tikzpicture}
\end{tabular}
\end{table}

Circuits using EML operator as a new element (Table~\ref{EML-symbol}) might be useful for analog computing \cite{Ulmann+2022}. One of the old problems in this field is construction of predefined multivariate elementary functions \cite{Ulmann03032024}. Using EML compiler (Sect.~\ref{subsec:compiler}), we can convert any expression (Fig.~\ref{fig:trees}) to such a circuit, with the topology of a binary tree. EML tree provides a uniform treatment of generic elementary functions.

\begin{table}[]
    
    \caption{
	Complexity of various functions in EML tree representation. EML Compiler column gives RPN code length $K$ for expressions generated from EML compiler. The value of $K$ of EML formula can be computed using e.g. Mathematica \texttt{LeafCount}. For the identity function $x$, the compiler returns $x$ directly (leaf count 1); the shortest non-trivial EML expression have leaf count 9. Last column show results of direct exhaustive search for shortest expressions. Numbers in parentheses show length of formulas which do not use the extended reals ($\pm$inf in floating-point). If search timed out, reached lower limit for $K$ is given. 
	\label{tbl:leaf_count}
	} 
    
    \centering
    \begin{tabular}{c|cc}
     Constant & EML Compiler & Direct search  \\
      \hline
        $1$              &  1           &  1 (1)  \\
		$0$              &  7           &  7 (7)  \\
        $-1$             &  17          & 15 (17) \\
        $2$              &  27          & 19 (19) \\
        $-2$             &  43          & 27 (27) \\
        $1/2$            &  91          & 29 (35) \\
        $-1/2$           &  107         & 31 (37) \\
        $2/3$            &  143         & 39 (39) \\
        $-2/3$           &  159         & 45 (47) \\
        $\sqrt{2}$       &  165         & $>$47 \\
        $i$              &  131         & $>$55  \\
        $e$              &  3           &  3  \\
        $\pi$            &  193         & $>$53  \\
    \end{tabular}

    \begin{tabular}{c|ccc}
      Function & EML Compiler & Direct search  \\
      \hline
        $x$              &  1           & 9    \\
        $e^x$            &  3           & 3    \\
        $\ln{x}$         &  7           & 7    \\ 
        $-x$             &  57          & 15   \\
        $\frac{1}{x}$    &  65          & 15   \\
        $x-1$            &  43          & 11   \\
        $x+1$            &  27          & 19   \\
        $x/2$            &  131         & 27     \\
        $2x$             &  131         & 19     \\
        $\sqrt{x}$       &  139         & 43 $\geq ? >$35  \\
		$x^2$            &  75          & 17   \\
    \end{tabular}
    \begin{tabular}{c|ccc}
     Operator & EML Compiler & Direct search \\
      \hline
        $x-y$              &  83         & 11 (11) \\
		$x+y$              &  27         & 19 (19) \\
        $x \times y$       &  41         & 17 (17) \\
        $x/y$              &  105        & 17 (17) \\
        $x^y$              &  49         & 25      \\
        $\log_x{y}$        &  117        & 29      \\
        $(x+y)/2$          &  287        & $>$27\\
        $x^2+y^2$          &  175        & $>$27\\
    \end{tabular}
    
\end{table}

\subsection{\label{subsect:continuos} Symbolic Regression by continuous optimization }

Modern symbolic regression (SR) methods \cite{udrescu2020ai,cranmer2020discovering,cranmer2023pysr} aim to discover closed-form expressions from data, but they typically search over heterogeneous grammars involving many distinct operators. Knowledge of a single operator, \eqref{eml}, which is sufficient to compute any elementary function, allows us to create a multiparameter ''master'' formula. Such a master formula, easily constructed in EML form due to binary expression tree, for some combinations of its parameters, is equivalent to elementary functions. By construction, such a general tree includes all possible formulas up to the given leaf count (tree depth). These trees are big by the standards of a typical mathematical analysis course, but small compared to what is in use in modern AI. For example, a full binary tree of depth $n$ has a total of $2^n-1$ branches and $2^n$ leaves. For the largest transformers with trillions ($10^{12}$) of parameters, the equivalent tree would have a depth of 40. The large values in Table~\ref{tbl:leaf_count} reflect the unoptimized prototype EML compiler (Subsect.~\ref{subsec:compiler}); direct exhaustive search yields substantially shorter expressions, as the rightmost column demonstrates.

The master formula can be constructed as follows. For simplicity we present the univariate case; the construction extends to an arbitrary number of input variables.
Both inputs to $\eml{(x,y)}$ can be: either 1, input variable $x$, or the result from the previous $\eml$, which we denote $f$. Let us represent such a general input by a linear combination
\begin{equation}
\label{3inputs}
\alpha_i + \beta_i x + \gamma_i f.    
\end{equation}
Now, setting specific values, we can recover all three cases
\begin{itemize}
\item 1, for $\alpha_i=1, \beta_i=0, \gamma_i=0$, 
\item x, for $\alpha_i=0, \beta_i=1, \gamma_i=0$, 
\item f, for $\alpha_i=0, \beta_i=0, \gamma_i=1$. 
\end{itemize}
At lowest tree level (leaves), there are only 1 and x, so there are no parameters $\gamma_i$.

As an example, I present full level-2 master formula
\begin{align*}
F(x) = \eml \biggl[
&\alpha_1 + \beta_1 x + \gamma_1  \eml{(\alpha_3 + \beta_3 x,\alpha_4 + \beta_4 x)},\\
&\alpha_2 + \beta_2 x + \gamma_2  \eml{(\alpha_5 + \beta_5 x,\alpha_6 + \beta_6 x)}
\biggr].
\end{align*}
If we set $\alpha_1=0, \beta_1=1, \gamma_1=0$ and $\alpha_2=1, \beta_2=\gamma_2=0$, we recover $\exp(x)$. 
Using $\alpha_1=\alpha_2=1$ and all $\beta_i=\gamma_i=0$ we get the constant $e$. Setting $\alpha_1=\beta_1=0, \gamma_1=1$, $\alpha_2=1, \beta_2=\gamma_2=0$, $\alpha_3=0, \beta_3=1, \gamma_3=0$ and $\alpha_4=1, \beta_4=\gamma_4=0$, we obtain double exponential $\exp{(e^x)}$. The total number of parameters for the level-2 master formula is 14. In general, the level-$n$ master formula has $5 \times 2^n-6$ parameters, see Supplementary Information, Part III.

While one could, in principle, create such a master formula using more usual elementary functions (+,-,/,**, $\sin, \cos$, \ldots, etc.), it would be ridiculously complicated and would lack any regular structure. Moreover, in practical SR one typically works with a reduced subset of operations, running the risk that the chosen set is incomplete, i.e., unable to express all elementary functions. The EML master formula is complete by design.

Number of parameters in \eqref{3inputs} can be reduced to two using some switch function. Or, alternatively, in a more modern approach, one can treat parameters $\alpha_i,\beta_i, \gamma_i$ as logits, and pass them through softmax function \cite{jang2017categorical} to convert them into probabilities normalized to
$\alpha_i+\beta_i+\gamma_i=1$. In two examples below, I'll use both methods.

The simplest proof of concept is provided by fitting some numerical data obtained from example elementary function,  the $\ln{x}$, using the complete level 3 binary tree with the above structure. Using simplex reparameterization, the number of free parameters reduces from 34 to 20. I managed to find all weights using plain \texttt{NMinimize}, a 'black-box' optimizer from Mathematica \cite{WolframMathematica}. The mean squared fitting error is at the level of numerical precision, and the resulting formula, after rounding weights to the nearest vertex of the simplex (i.e.\ snapping each to~0 or~1), is \textbf{exactly} $\ln{x}$. Not only in the provided data range, but also beyond that. Generalization/extrapolation is nearly perfect in the above example. See Wolfram Mathematica notebook \texttt{Log\_fit.nb} from EML toolkit \cite{SymbolicRegressionPackage} and Supplementary Information (Sect III.B) for details.

In practice, such a naive one-liner approach does not scale. Therefore, I performed several experiments using more recent techniques of Machine Learning. Simple functions of two variables, taken from composed EML, such as $\ln(e-\ln(e^x - \ln{y}))$, were used as targets.  Python code was created, which used \texttt{DTYPE = torch.complex128} data type. Main issues encountered during training of EML net were related to range overflow due to multiply composed exponentials, as well as floating-point errors (NaN) caused by the specific implementation of complex arithmetic. They can be eliminated by clamping arguments and values for $\exp{(x)}$, and careful (i.e. without impairing torch automatic differentiation) inspection of both real and imaginary parts. Parameters from  \eqref{3inputs} were treated as logits. Optimization was a multi-stage process: first, usual training using a stochastic gradient optimizer (Adam), then a hardening phase pushing weights in \eqref{3inputs} towards 0 or 1. Finally, weights were clamped to exact symbolic values. Systematic experiments (over 1000 runs with varied seeds and initialization strategies) show that blind recovery from random initialization succeeds in 100\% of runs at depth~2, approximately 25\% at depths 3--4, and below 1\% at depth~5. At depth~6, no blind recovery was observed in 448 attempts. When successful, the snapped weights yield mean squared errors at the level of machine epsilon squared ($\sim 10^{-32}$), consistent with exact symbolic recovery. Noteworthy, when the weights of the correct EML tree are perturbed by Gaussian noise, the optimization converges back to the exact values in 100\% of runs, even for trees of depth~5 and~6. This demonstrates that the EML tree representation is valid and that the correct basins of attraction exist. Finding them from random initialization becomes harder with depth. If one manages to improve convergence of EML trees beyond proof of concept, possibly using another binary operator similar to \eqref{Sheffers} but with better properties (non-exponential asymptotics, no domain issues), then we could achieve symbolic regression of data using gradient-based methods \cite{petersen2021deep}. See Supplementary Information for systematic training experiments, code and details.

\section{Conclusions and open questions \label{sec:conclusions}}

The operator EML, \eqref{eml}, provides a single sufficient primitive from which real elementary functions can be constructed and evaluated.
Consequently, a wide class of computations built from such functions can also be cast in EML form. 
It is not unique; several close variants are likely to exhibit similar properties, including EDL, \eqref{edl-e}, and the swapped-argument form $-\eml(y,x)$, \eqref{eml-infty}.
 More operators of this kind exist. Perhaps an entire continuous family of them awaits discovery, with properties more convenient than \eqref{eml}. For example, the requirement for one of the constants: $-\infty,1,e, \ldots$ to be always present among terminal symbols makes its use less elegant and more complicated (cf.~Subsect.~\ref{subsect:continuos}) in comparison with, e.g., standard neural nets or the NAND gate. The latter is able to generate\footnote{
E.g., $1 \equiv \text{NAND}(x,\text{NAND}(x,x))$. In practice, one will rarely use this property. Usually, we use directly available 0s and 1s, e.g. GND and $V_{cc}$ in digital circuits.} 
0s and 1s out of `anything'. The EML operator does not have this useful property. Whether an EML-type binary Sheffer working without pairing with a distinguished constant exists is an open question. Proving such impossibility for any given candidate is non-trivial: one might expect $f(x,x)$ being constant to suffice, but consider $B(x,y) \! = \!  x \!-\! \tfrac{y}{2}$, for which $B(x,x) \! = \!  \tfrac{x}{2}$ yet $B(B(x,x),x) \! = \! 0$. Such traps illustrate why systematic search is essential in this work. A {\em ternary} operator, $T(x,y,z)\! = \! {e^x}/{\ln{x}} \times {\ln{z}}/{e^y}$, for which $T(x,x,x)\! =\! 1$ is next candidate for further analysis \cite{AO-in-preparation}.

Whether a univariate Sheffer exists, serving simultaneously as a neural activation function \cite{DubeyActivationSurvey2022} and as a generator of all elementary functions, remains open (see SI, Sect.~5).

One might complain that the EML representation of elementary functions requires complex arithmetic for real math, at least internally. 
Just as quantum computing uses complex amplitudes to compute real probabilities, EML uses complex intermediates to compute real elementary functions. This seems inevitable. We must somehow compute the imaginary unit $i$, $\pi$, and all trigonometric/hyperbolic functions \emph{via} Euler's formula, \eqref{Euler}. For that, we use $\ln{x}$ for $x<0$. A continuous Sheffer working purely in the real domain seems impossible. My search for alternatives, e.g., using pairs of trigonometric/hyperbolic functions and their inverses instead of $\exp/\ln$, found nothing.  Quite surprisingly, the requirement to use complex numbers internally causes only minor problems in practice of using \eqref{eml} in Computer Algebra Systems or numerical simulations.

Since standard activation functions are themselves elementary, any conventional neural network is a special case of an EML tree architecture. Current networks can learn symbolic algebra \cite{lample2020deep} and digit-level arithmetic \cite{lee2024teaching}, but their internal mechanisms remain opaque \cite{liu2025kan}, and efficient exact evaluation of elementary functions as continuous real-valued operations is still beyond their reach. EML representations go further: as demonstrated in Subsect.~\ref{subsect:continuos}, trained weights can snap to exact binary values, recovering closed-form elementary subexpressions alongside approximations. When this succeeds, the discovered circuits are legible as elementary function expressions --- a form of interpretability unavailable to conventional architectures.

\subsection*{AI use disclosure}
The core idea, the discovery of the EML Sheffer operator, the verification methodology, and results are entirely the author's own work.
Large language models (including recent Claude, Grok, Gemini and ChatGPT) were used mainly for language editing and coding assistance. 

\subsection*{Data availability}
Code, scripts, and supplementary reproducibility materials used to generate the figures, tables, and numerical results are available in the SymbolicRegressionPackage repository~\cite{SymbolicRegressionPackage}. An archival snapshot of the exact submission version is deposited in Zenodo DOI: 10.5281/zenodo.19183008 (\url{https://doi.org/10.5281/zenodo.19183008}). The archival package contains the source code, scripts, and README files needed to rerun the reported results.

\subsection*{Acknowledgments}
Computational resources were partially provided by Google Cloud Research Credits.

\subsection*{Supplementary Information}

Extensive three-part Supporting Information is supplied as a separate SI Appendix PDF. 

\bibliographystyle{unsrtnat}
\bibliography{EML}

\end{document}